\documentclass{aastex631}

\usepackage{amsthm,amsmath,amssymb}
\usepackage{lipsum}
\usepackage{float}
\usepackage{soul}

\begin{document}

\title{Asteroid Rotation Periods: Statistical Analysis in the Diameter-Spin Distribution}

\author[0009-0006-9150-3392]{Maryam Nastaran}
\affiliation{Faculty of Geology, University of Tehran, 1417935840 Tehran, Iran}

\author[0000-0002-0196-9732]{Atila Poro}
\altaffiliation{iotamiddleeast@yahoo.com}
\affiliation{The International Occultation Timing Association Middle East Section, 15875 Tehran, Iran}
\affiliation{LUX, Observatoire de Paris, CNRS, PSL, 61 Avenue de l'Observatoire, 75014 Paris, France}
\affiliation{Astronomy Department of the Raderon AI Lab., BC., Burnaby, Canada}

\author[0000-0002-7041-0705]{Raziyeh Hosseini}
\affil{Department of Earth Sciences and Engineering, Missouri University of Science and Technology, Rolla, MO 65409, USA}

\author[0009-0003-8294-6296]{Matin Najarzadeh}
\affiliation{The International Occultation Timing Association Middle East Section, 15875 Tehran, Iran}
\affil{University of Tehran, 1417935840 Tehran, Iran}

\begin{abstract}
This study examines the rotational characteristics of asteroids through statistical modeling of the diameter–period relationship. A statistical evaluation of the diameter–period relationship was conducted using a dataset of 34,326 asteroids. Clustering identified three main groups, including a dense cluster below the spin barrier, a population of small, fast-rotating asteroids, and a more diffuse group. Geometric and density-based analyses showed that the densest region consists of objects with diameters from 3 to 10 km and rotation periods between 3 and 9 hours, some of which extend beyond the spin barrier. Polynomial modeling demonstrated that a third-degree fit provides the most stable representation of the overall trend without overfitting. Additionally, an empirical lower boundary was identified and proposed, below which no asteroid was found in either the main sample or the selected targets.
\end{abstract}


\section{Introduction}
Asteroids originated from remnants of the protoplanetary disk surrounding the Sun (\citealt{muinonen2022asteroid}, \citealt{vavilov2025rotation}, \citealt{johansen2015growth}). The composition, size, and orbital distribution of asteroids provide evidence of past events in the solar system (\citealt{demeo2014solar}, \citealt{morbidelli2015dynamical}, \citealt{clement2020record}), although their current population exhibits a distribution different from that of the early solar system (\citealt{vavilov2025rotation}). This transformation, driven by billions of years of collisions and fragmentation, has led to an increase in their number and the formation of asteroid families (\citealt{hirayama1918groups}, \citealt{zappala1990asteroid}, \citealt{milani2014asteroid}).

Studying the dynamics of asteroids provides valuable insights into the characterization of the numerous planetary systems discovered to date (\citealt{warner2011using}). In particular, determining and analyzing their rotation rates is a fundamental aspect of asteroid research and is essential for understanding their evolution both as individual bodies and as integral components of the solar system (\citealt{warner2011using}, \citealt{2016IAUS..318..177D}).

Photometric data, widely used in asteroid research, provide an effective means for continuously investigating their physical properties and rotation periods (\citealt{vdurech2022rotation}; \citealt{warner2011using}). Moreover, studies such as \cite{vdurech2009asteroid}, \cite{cellino2009genetic}, \cite{hanuvs2011study}, and \cite{hanuvs2013asteroids} have integrated photometric data with other techniques to develop more comprehensive asteroid models.

Ongoing research into the relationship between asteroid diameter and rotational speed offers fascinating insights. The spin barrier for asteroids was first proposed by \citet{harris1996rotation} in 1996, identifying a 2.2-hour rotation limit for asteroids with diameters $\gtrsim 1$ km. This limit reflects the gravitational constraints of larger asteroids with rubble-pile structures. \citet{pravec2000fast} observed that asteroids larger than a few hundred meters typically rotate below this threshold, supporting the rubble-pile model. Further, \citet{pravec2006nea} demonstrated that asteroids with diameters greater than 200 m generally cannot rotate faster than this critical period due to insufficient tensile strength. In contrast, smaller asteroids can spin more rapidly and are more likely to be monolithic \citep{chang2014313}. Rare exceptions to the spin barrier include asteroid 2001 OE84, with a diameter of 0.9 km and a rotation period of 29.19 minutes \citep{2002ESASP.500..743P}, and asteroid (68063) 2000 YJ66, which rotates near the 2.2-hour threshold \citep{warner2015trio}. Another notable exception is asteroid (335433) 2005 UW163, with a rotation period of 1.290 hours \citep{chang2014313}. However, given that the spin barrier is influenced by both size and composition, it is reasonable to expect that larger asteroids are more prone to structural failure at high rotation rates \citep{pravec1998lightcurves}.

The aim of this study is to explore the statistical structure of the diameter–rotation period relationship among asteroids and to uncover underlying patterns in their rotational behavior. By utilizing a large dataset and applying both statistical and geometric modeling techniques, this work seeks to provide a more comprehensive understanding of how asteroids are distributed in diameter–period space. Additionally, the study aims to establish empirical constraints on asteroid rotation and to develop a stable model that captures the overall trend of this relationship.

\vspace{0.4cm}
\section{Diameter-rotation period relationship}
Extensive research has been devoted to understanding the relationship between asteroid rotational speed and diameter, an area that continues to grow with the increasing number of discovered asteroids and advancements in data analysis techniques. Observational data reveal that rotation periods span a broad range, from a few minutes to several days, although most fall within the 2 to 20 hour interval. A notable feature in this distribution is the marked absence of asteroids larger than approximately 200 meters with rotation periods shorter than 2.2 hours (\citealt{pravec2006nea}). In contrast, smaller asteroids, particularly those with diameters under approximately 150 meters, are occasionally observed to exceed this threshold.

We analyzed the distributional structure of asteroid rotation periods as a function of diameter using a dataset comprising 34,326 asteroids from the Asteroid Lightcurve Data Base (LCDB; \citealt{2021Warner})\footnote{Revised version October 2023}. The majority of these asteroids belong to the main belt, while the sample also includes a small fraction of objects from other regions. This sample provides a comprehensive representation of the diverse asteroid population and enables a large-scale statistical investigation of rotational behavior. To prepare the data for analysis, both diameter and rotation period values were logarithmically transformed to improve scaling and compatibility with distance-based analytical methods.

\begin{figure*}
\centering
\includegraphics[width=0.477\textwidth]{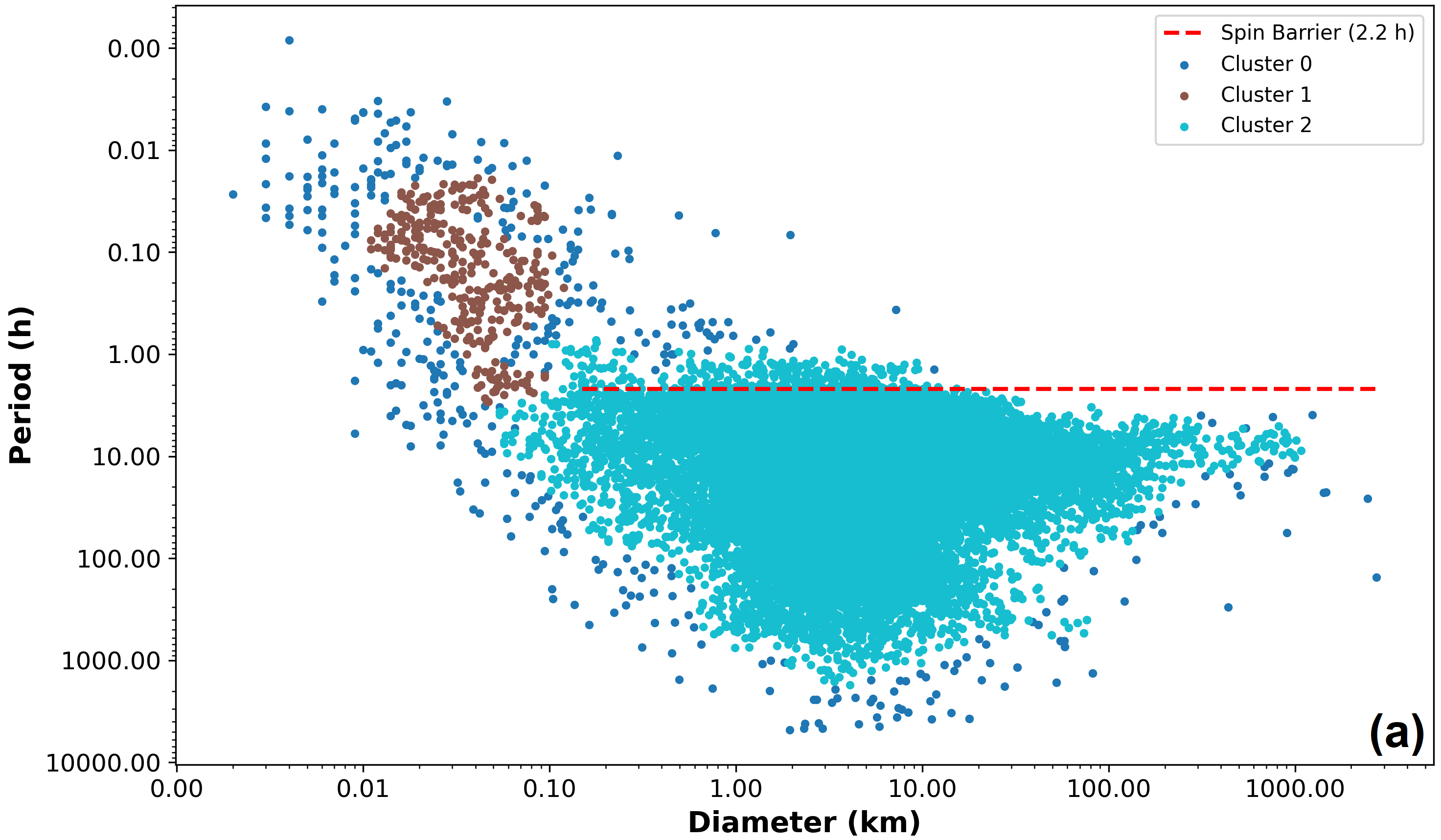}
\includegraphics[width=0.477\textwidth]{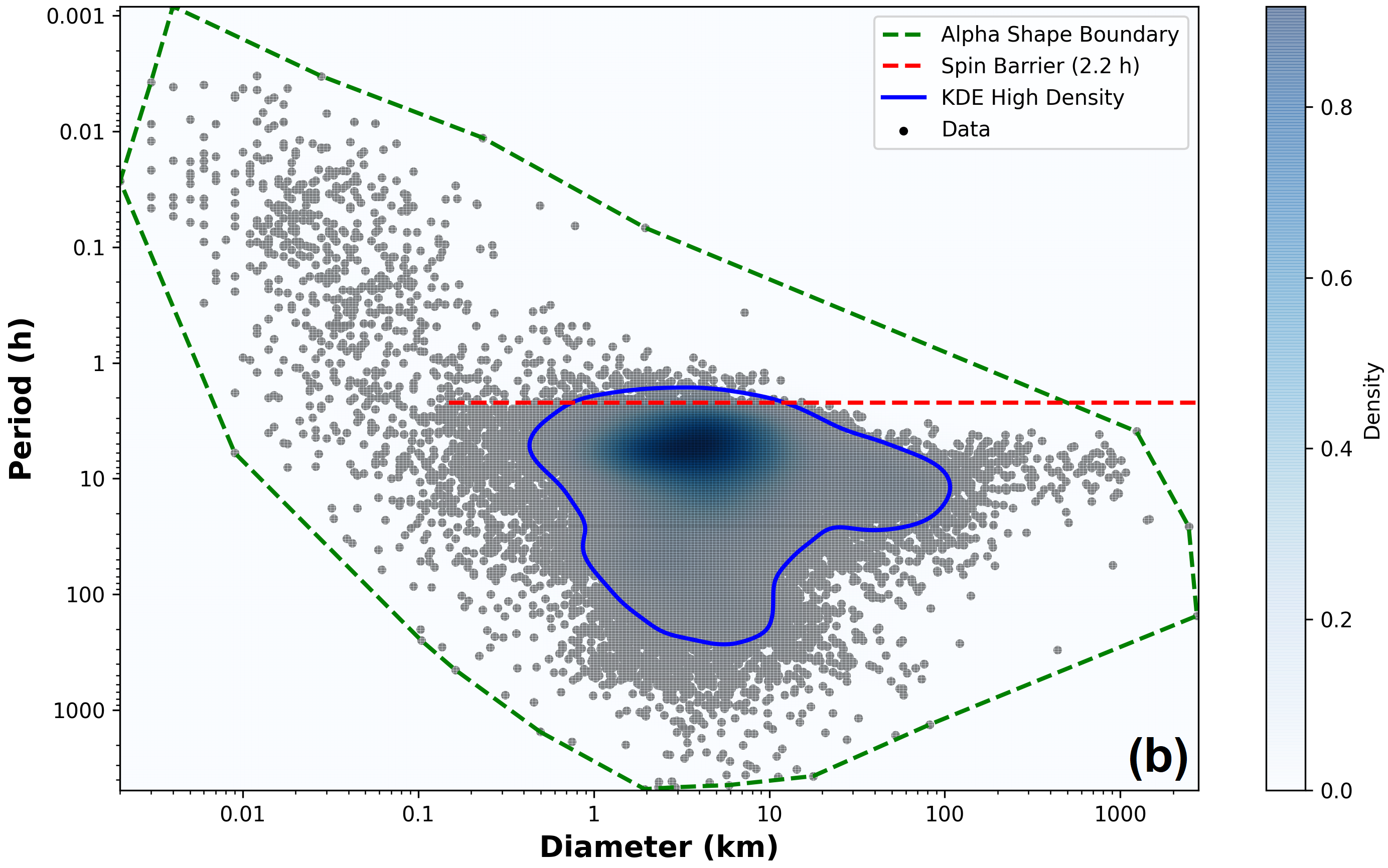}
\includegraphics[width=0.477\textwidth]{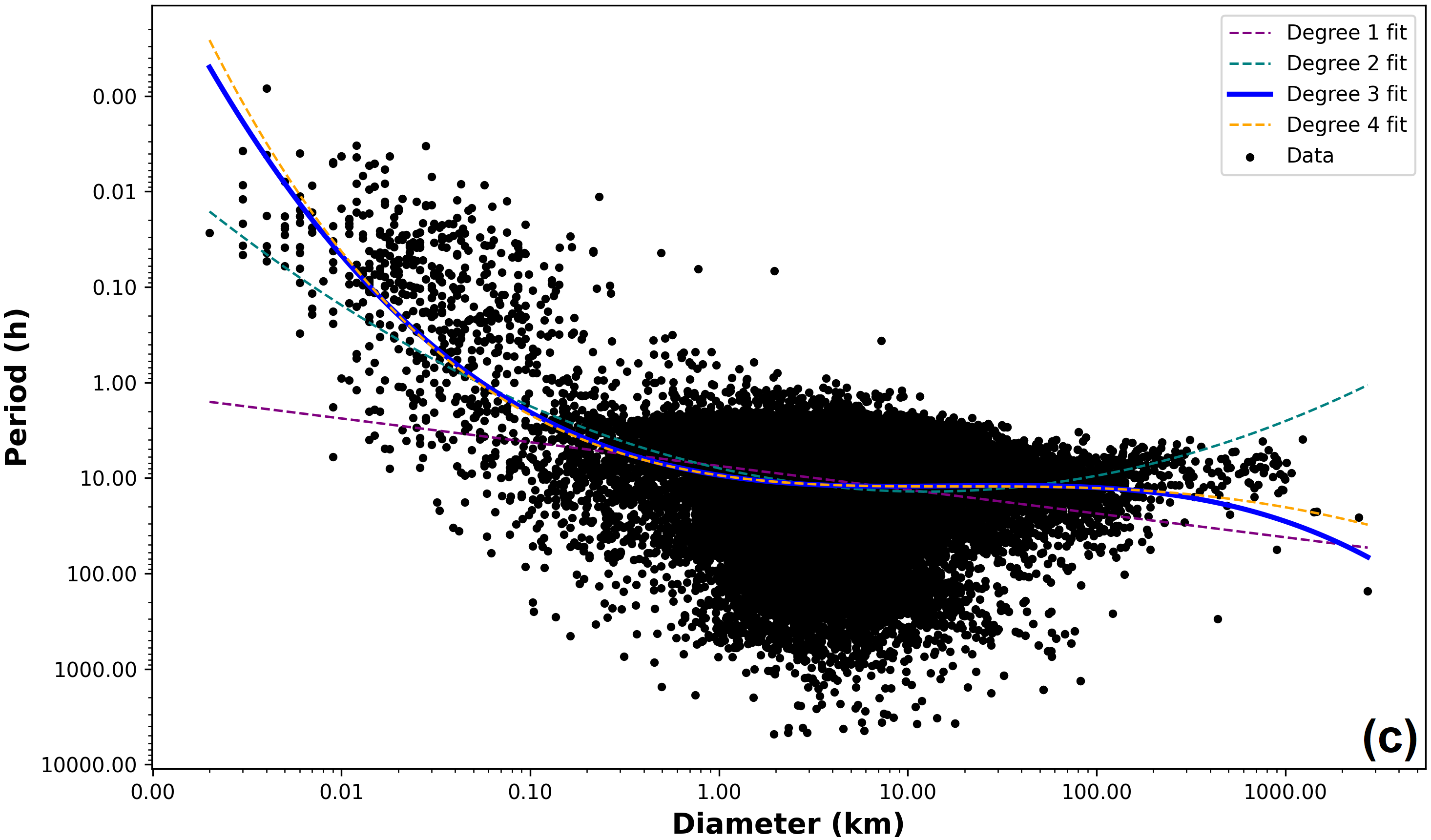}
\includegraphics[width=0.477\textwidth]{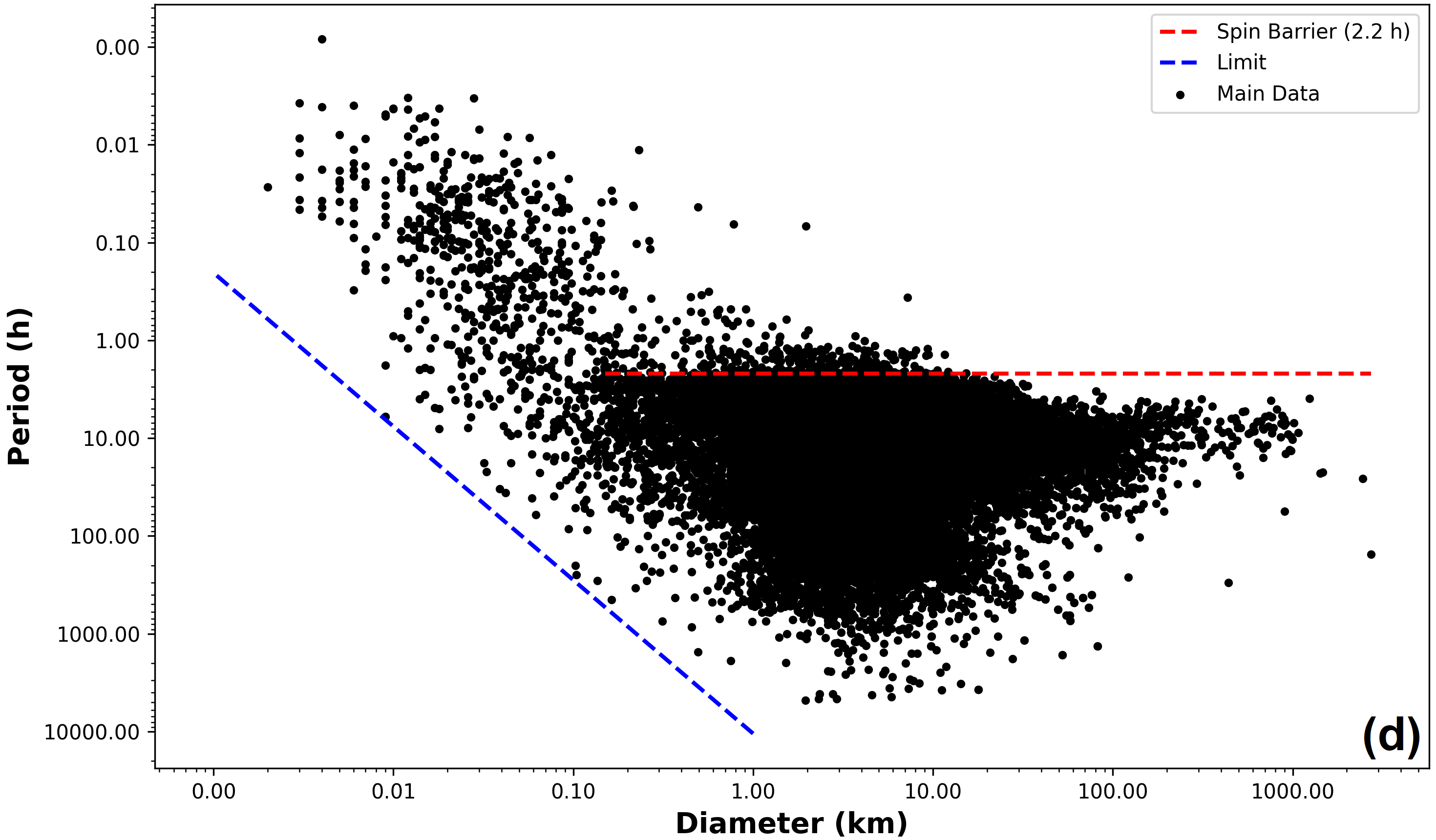}
\caption{Distribution and structural analysis of asteroid rotation periods as a function of diameter in logarithmic space.
(a) DBSCAN clustering results, showing three main groups along with the 2.2-hour spin barrier (red dashed line).
(b) Combined visualization of geometric boundary (alpha shape), density structure (KDE heatmap), and high-density region (blue contour).
(c) Polynomial fits (degrees 1-4) modeling the rotation–diameter trend.
(d) Blue dashed line marking an empirically derived lower boundary.}
\label{Fig2}
\end{figure*}

To identify intrinsic structural patterns within the data, the Density-Based Spatial Clustering of Applications with Noise (DBSCAN; \citealt{1996kddm.conf..226E}) algorithm was employed. Due to its ability to detect clusters of arbitrary shape and size without requiring \textit{a priori} specification of the number of clusters, DBSCAN is particularly well-suited for astronomical datasets characterized by heterogeneous densities and inherent noise. Prior to clustering, the data were normalized using standardization (StandardScaler), ensuring that both log-diameter and log-period features had zero mean and unit variance. The DBSCAN parameters were selected empirically (\texttt{eps} = 0.2, \texttt{min\_samples} = 10) to achieve a balance between cluster separability and structural stability. The initial clustering output included a number of small clusters (fewer than 50 asteroids), which were statistically unreliable. To mitigate the risk of misinterpretation, members of these low-density clusters were reassigned to the nearest larger cluster based on Euclidean distance in the normalized feature space. The final results of this clustering procedure are presented in Figure \ref{Fig2}a, where each cluster is rendered in a distinct color. Additionally, a red dashed line at a constant rotation period of 2.2 hours is included to indicate the spin barrier (\citealt{pravec2006nea}).
As shown in Figure \ref{Fig2}a, two main clusters were identified: the largest and densest cluster lies below the spin barrier line, and the second cluster consists of small asteroids with high rotation speeds. A third, more scattered group is also visible across the diagram. In the DBSCAN output, these objects are labeled as cluster 0 and formally treated as noise, although their distribution may still reflect underlying physical processes rather than purely random scatter.
\\
\\
Further examination of the boundary and density distribution of asteroids in the log-diameter–log-period space was conducted using two complementary techniques. The alpha shape method (\citealt{edelsbrunner1983shape}) was used to delineate the outer boundary, while Kernel Density Estimation (KDE; \citealt{silverman2018density}) was applied to characterize the internal density structure.

The alpha shape was constructed using an empirically chosen parameter of $\alpha=0.12$, allowing for a more precise modeling of the dataset’s outer envelope compared to the convex hull. This method is capable of capturing non-convex and complex geometries, making it particularly suitable for delineating the occupied region of scattered data. In parallel, KDE with a bandwidth of 0.2 was employed to estimate the probability density over the two-dimensional domain. The resulting density field was visualized as a heatmap, with color intensity indicating the relative concentration of points. In addition, a continuous contour line was drawn at a high-density threshold (corresponding to 8\% of the maximum density) to highlight the densest subregions. The combined visualization is presented in Figure \ref{Fig2}b, where the raw data points, geometric boundary, density distribution, and high-density regions are simultaneously displayed. This integrated representation provides a comprehensive view of the spatial structure and serves as a robust foundation for subsequent analyses. One notable observation in Figure \ref{Fig2}b is that part of the dense region, outlined by the blue line, extends above the spin barrier boundary.
\\
\\
To analyze the general trend in the distribution of asteroid rotation periods with respect to diameter, polynomial fitting was performed in the logarithmic diameter–period space. Polynomial models of degrees 1 through 4 were fitted to the log-transformed data using the least squares method, allowing the relationship to be modeled at different levels of complexity. To evaluate the performance of each model, the root mean square error (RMSE; \citealt{feigelson2012modern}) was computed for every fit. The RMSE is defined as:
\begin{eqnarray}
\mathrm{RMSE} = \sqrt{\frac{1}{n} \sum_{i=1}^{n} \left(y_i - \hat{y}_i\right)^2}
\end{eqnarray}
\noindent where \( y_i \) represents the observed values (log-transformed periods), and \( \hat{y}_i \) denotes the predicted values from the fitted model. The results showed that the RMSE values for degrees 1 to 4 were 0.5626, 0.5418, 0.5366, and 0.5365, respectively. The marginal improvement in RMSE from degree 3 to 4 indicates that increasing the model complexity beyond degree 3 yields negligible gains in accuracy. Furthermore, the degree-4 fit exhibited signs of overfitting, particularly in regions with low data density. In addition to RMSE, the residuals of each model were examined to detect any systematic patterns or non-random deviations. This analysis revealed that the degree-3 fit provided a favorable balance between prediction accuracy and homogeneity in residual distribution. Accordingly, the degree-3 polynomial was selected as the final model, as it achieved a satisfactory RMSE of 0.5366 while maintaining simplicity and robustness relative to more complex alternatives. The fitted curves are illustrated in Figure \ref{Fig2}c.
\\
\\
A blue dashed line was added to the plot to indicate a boundary we identified empirically, below which no asteroid data are present in the entire sample (Figure \ref{Fig2}d). This absence may indicate the existence of a physical or dynamical constraint that limits the rotational behavior of small asteroids, preventing rotation periods beyond a certain threshold.

\vspace{0.4cm}
\section{Discussion and Conclusion}
The investigation of asteroid rotation periods as a function of diameter revealed several consistent and interpretable patterns:

$\bullet$ Clustering analysis identified three well-separated groups of objects in the log-diameter–log-period space. The largest and densest group is concentrated mostly below the 2.2-hour spin barrier, aligning with the theoretical limit for gravitationally bound, low-cohesion asteroids (\citealt{pravec2008spin}). Nonetheless, a fraction of this group appears above the boundary, suggesting a gradual rather than abrupt transition. A second, smaller group consists of fast-rotating asteroids with diameters typically less than 200 meters. The third group includes more dispersed objects with no dominant trend.

$\bullet$ Geometric boundary extraction using the alpha shape method, together with kernel density estimation, provided a detailed representation of the data’s spatial structure. The densest concentration of objects was found among asteroids with diameters between 3 and 10 kilometers and rotation periods between 3 and 9 hours. Notably, a portion of this high-density region extends above the 2.2-hour threshold, which may be attributed to measurement uncertainties. These findings suggest that the current understanding of the spin barrier in asteroids may require reevaluation.

$\bullet$ Polynomial modeling of the diameter–period relationship showed that a third-degree polynomial offered an optimal balance between model accuracy and complexity. Higher-order models did not significantly improve the fit and introduced overfitting, especially in sparsely populated regions. Residual analysis confirmed that the cubic model was sufficiently robust for representing the overall trend.

It should be noted that in our clustering analysis, only two physical parameters, diameter and rotation period, were used. The subsequent polynomial fitting was therefore applied as a descriptive tool to outline the empirical boundary of the distribution rather than as a predictive model. Thus, the present results should be regarded as a statistical characterization of existing data rather than a classification tool for newly discovered asteroids. While in principle DBSCAN could be applied to new data if diameter and period are known, the current framework cannot assign objects based on other features that were not included in the clustering.

$\bullet$ An empirical lower boundary was identified, beneath which no asteroid is present from the dataset. The absence of objects in this region may reflect an inherent constraint on the spin behavior of small asteroids, potentially imposed by internal cohesion, collisional evolution, or observational limitations.

$\bullet$ The observed presence of several asteroids above the nominal spin barrier may be linked to the Yarkovsky-O'Keefe-Radzievskii-Paddack (YORP; \citealt{bottke2006yarkovsky}) effect. This torque, arising from the anisotropic absorption and re-emission of solar radiation, can gradually alter the spin states of small bodies over timescales of $10^{5}$-$10^{7}$ years \citep{rubincam2000radiative,bottke2006yarkovsky}. In particular, the YORP effect is capable of both accelerating and decelerating rotation rates, potentially driving some small asteroids beyond the expected spin-barrier threshold. Therefore, the apparent outliers above the barrier in our analysis may not solely represent noise in the data, but could also reflect evolutionary processes induced by YORP. Future studies that couple rotational statistics with dynamical and thermal modeling may help to disentangle the role of YORP in shaping the diameter-period distribution.

\clearpage
\section*{Acknowledgments}
This manuscript was provided by the International Occultation Timing Association Middle East Section (IOTA/ME, {\url{https://iota-me.com}}).

\bibliography{References}{}
\bibliographystyle{aasjournal}

\end{document}